\documentclass{iau}

\usepackage{amsmath}
\usepackage{graphicx}
\usepackage{multirow}
\newcommand{\apj}{{ApJ}}
\newcommand{\apjl}{{ApJL}}
\newcommand{\aj}{{AJ}} 
\newcommand{\mnras}{{MNRAS}}
\newcommand{\aap}{{Astronomy \& Astrophysics}}  

\begin{document}

\lefttitle{Das et al.}
\righttitle{IAU Symposium 379: Template}

\jnlPage{1}{7}
\jnlDoiYr{2023}
\doival{10.1017/xxxxx}

\aopheadtitle{Proceedings of IAU Symposium 379}
\editors{P. Bonifacio,  M.-R. Cioni \& F. Hammer, eds.}

\title{The oblateness of dark matter halos of nearby galaxies and its correlation with gas mass fractions}

\author{Mousumi Das$^{1}$, Roger Ianjamasimanana$^{2}$, Stacy McGaugh$^{3}$, James Schombert$^{4}$, K.S.Dwarakanath$^{5}$}
\affiliation{$^1$ Indian Institute of Astrophysics, Bengaluru 560034, India\\
$^2$ Instituto de Astrofísica de Andalucía (CSIC), Glorieta de la Astronomía, E-18008 Granada, Spain  \\
$^3$ Department of Astronomy, Case Western Reserve University, 10900 Euclid Avenue, Cleveland, OH 44106, USA\\
$^4$ Department of Physics, University of Oregon, 120 Willamette Hall, 1371 E 13th Avenue, Eugene, OR 97403461, USA\\
$^5$ Raman Research Institute, C.V. Raman Avenue, Sadashivanagar, Bengaluru, Karnataka 560080, India} 

\begin{abstract}
We present a method to measure the the oblateness parameter $q$ of the dark matter halo of gas rich galaxies that have extended HI disks. We have applied our model to a sample of 20 nearby galaxies that are gas rich and close to face-on, of which 6 are large disk galaxies, 8 have moderate stellar masses and 6 are low surface brightness (LSB) dwarf galaxies. We have used the stacked HI velocity dispersion and HI surface densities to derive $q$ in the outer disk regions. Our most important result is that gas dominated galaxies (such as LSB dwarfs) that have M(gas)/M(baryons)$>$0.5 have oblate halos (q$<0$.55), whereas stellar dominated galaxies have a range of q values from 0.2 to 1.3. We also find a significant positive correlation between q and stellar mass, which indicates that galaxies with massive stellar disks have a higher probability of having halos that are spherical or slightly prolate, whereas low mass galaxies preferably have oblate halos. We briefly also discuss how the halo shape affects the disks of galaxies, especially the oblate halos.
\end{abstract}


\maketitle

\section{Introduction}
Galaxy disks are known to be embedded in large dark matter halos, the main evidence being the flat rotation curves of disk galaxies \citep{kent1987}. The halos play a dominant role in galaxy evolution, as they affect disk stability and hence star formation \citep{yurin.springel.2015,sellwood.2016}. Although we know that halos are important for disk evolution and galaxy rotation curves indicate that they contain most of the galaxy mass, their detailed properties are not well constrained. Halo angular momentum or spin, halo shape or even halo masses are not well understood. We are also not certain where the halos really ends, as there is no clear tracer beyond the outer limits of the HI rotation curve. In this paper we focus on the determination of halo shape. We present a new method to determine the shape of dark matter halos using the neutral hydrogen (HI) velocity dispersion measured in face-on galaxies.

\begin{figure}[h]
\centering
  \includegraphics[scale=.27]{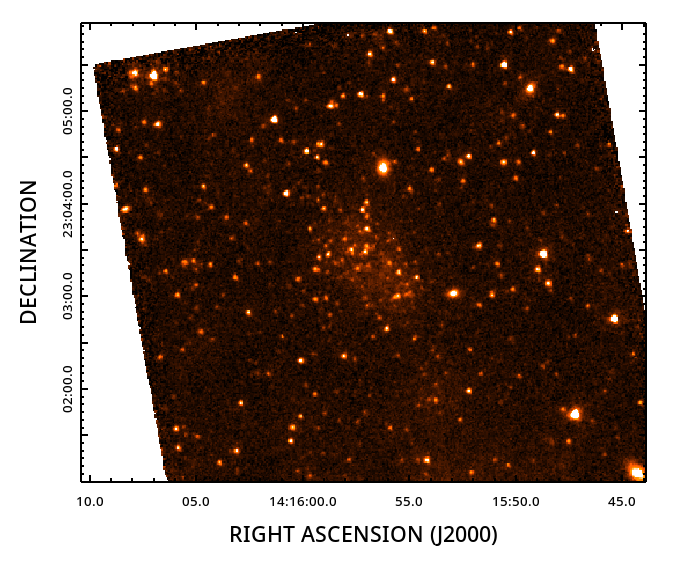}
  \includegraphics[scale=.27]{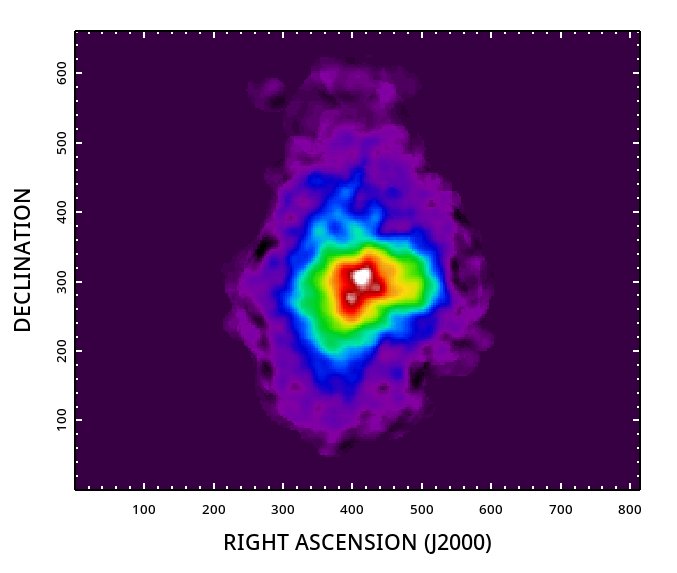}
  \caption{Comparison of the stellar and gas distribution in the dwarf galaxy DDO187. The image on the left is the Spitzer 3.6$\mu$m image and that on the right is the THINGS VLA HI 21cm emission map. Both Right Ascension (RA) and Declination are in J2000, and the images are matched in WCS. Note that the HI is more extended than the stellar disk as traced by the 3.6$\mu$m emission.}
\end{figure}

The earliest observational studies of halo shapes  used the polar rings around galaxies to model halo shapes \citep{combes2014}. But such rings are quite rare and so the method cannot be applied to a large sample of galaxies. Later studies used the flaring of HI disks in edge-on galaxies \citep{banerjee.jog.2008}, the kinematics of stars in our Milky way and other edge on galaxies \citep{olling1996}, the distribution of globular clusters in galaxies \citep{posti.helmi.2019} and more  recently the distribution and kinematics of stellar streams around galaxies \citep{pearson.etal.2022}.

Numerical simulations have also helped our understanding of halo shapes. Early studies showed that the gas infall, star formation and the resulting build-up of stellar mass within galaxy halos affects the triaxial nature of halos and makes them rounder \citep{dubinski1994}. This is also seen in recent cosmological simulations where the halo axes ratio in the disk plane (b/a) is often close to 1 \citep{prada.etal.2019}. The halos vary from prolate to oblate, but the Milky Way mass halos are close to spherical \citep{chua.etal.2019}. However, one disadvantage of these studies is that they do not resolve the smaller dwarf galaxies with stellar masses $<10^{8}M_{\odot}$. So the variation of halo shape with stellar or halo masses is not not well understood. 

In the following sections we briefly present a new way to determine the shapes of galaxy halos using the HI velocity dispersion in their extended disks. As is well known, the dark matter halo dominates the mass in the outer disks of galaxies, so our method focuses on the outer disk regions, where the stellar disk mass declines. Our method is applied to face-on disk galaxies rather than edge-on galaxies, as
previously investigated in the literature \citep{banerjee.jog.2011}. Our basic method is described in \citet{das.etal.2020}, where it has been shown that halo dark matter is important for the vertical support of extended HI disks and the method has been applied to a galaxy sample in \citet{das.etal.2023}. In this paper we first describe the need for halo dark matter for the vertical support of extended gas disks, the method to determine the halo vertical axis ratio $q$ (where $q=c/a$) and then its importance for understanding disk dynamics in galaxies.  

\section{The vertical support of galaxy disks}
Galaxy disks are supported by rotation in the radial direction where the gravitational force at any disk radius is balanced by the centrifugal force, so that v$^2$=GM(dyn)/R. This simple equality can be used to determine approximately the total mass within a radius R in a galaxy, but holds only if the dark matter halo is spherical. The vertical equilibrium in the disk is different. The gravitational binding force provided by the disk stars is balanced by the pressure in the gas and the velocity dispersion of the stars. As shown in \citet{das.etal.2020}, the disk may also contain a significant amount of dark matter that is generally associated with the halo. 

In the outer parts of galaxy disks where there is often HI gas but very little stellar mass, the role of the dark matter becomes much more important for maintaining the vertical disk equilibrium. Some good examples are galaxies such as NGC628 which has a very extended HI disk (see Das et al. 2020 for the image). Also, dwarf galaxies such as DDO187 (Figure 1) where the comparison of the near-infrared (NIR) and HI image clearly shows that the stellar disk is much less extended. Extreme examples of this are the LSB dwarfs lying in isolated voids such as KK246 \citep{kreckel.etal.2011} and dwarfs lying along cosmic filaments such as NGC4701 \citep{das.etal.2019}. 

\begin{figure}[h]
\centering
  \includegraphics[scale=.1]{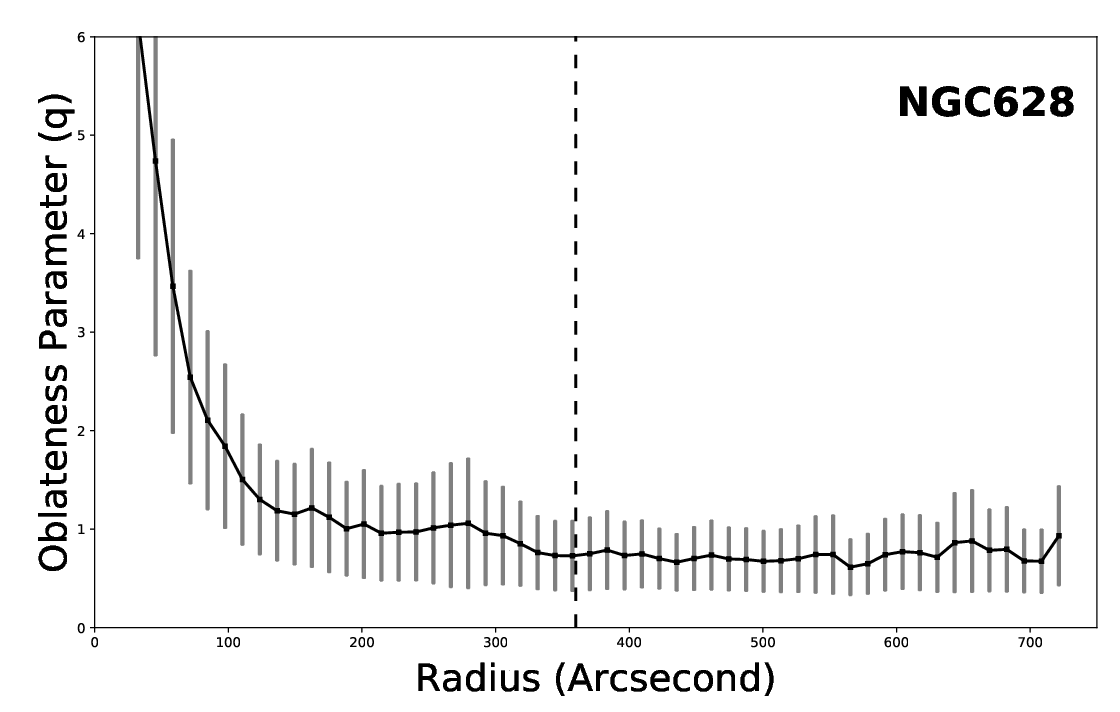}
  \includegraphics[scale=.1]{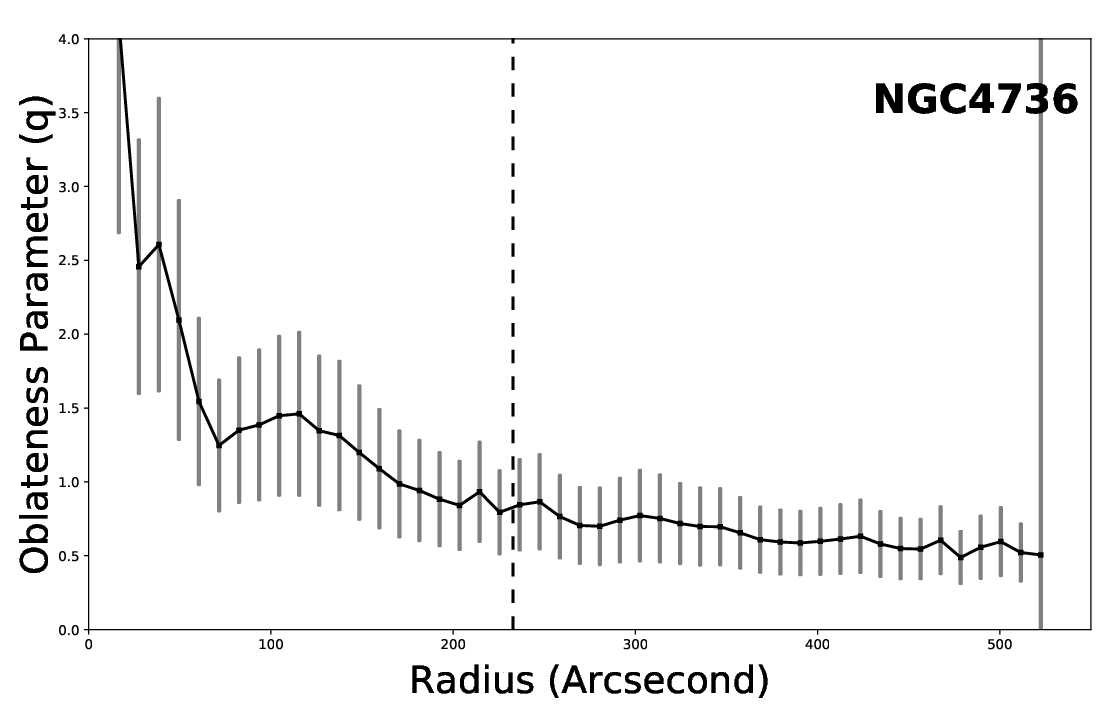}
  \includegraphics[scale=.1]{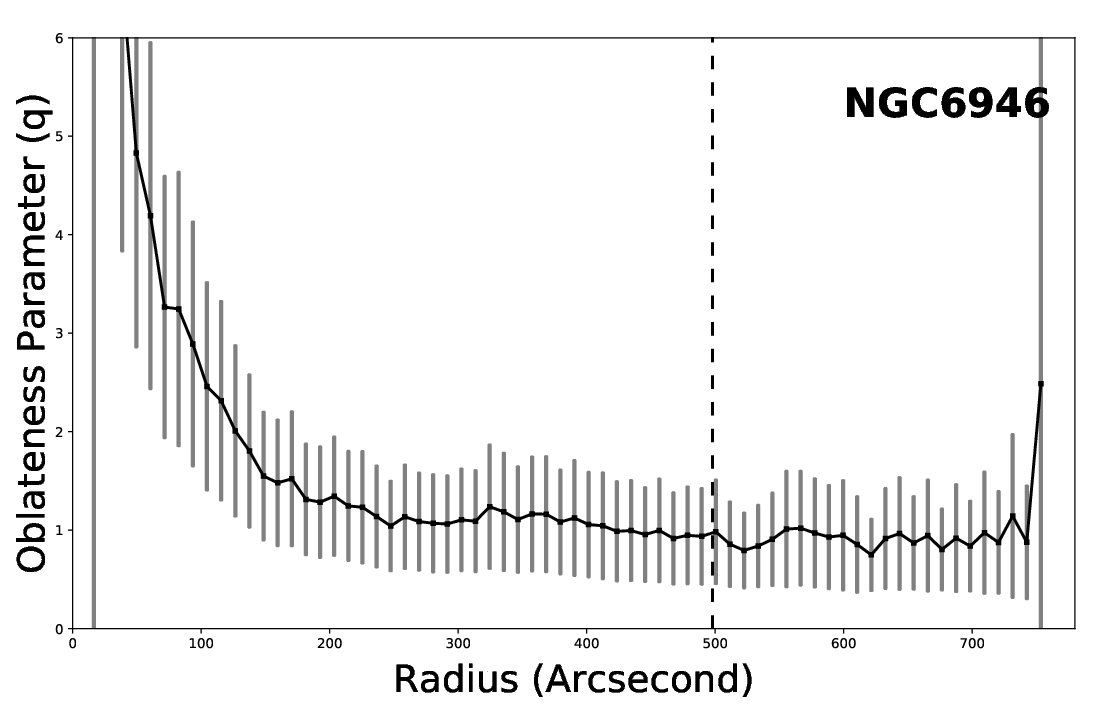}
  \includegraphics[scale=.1]{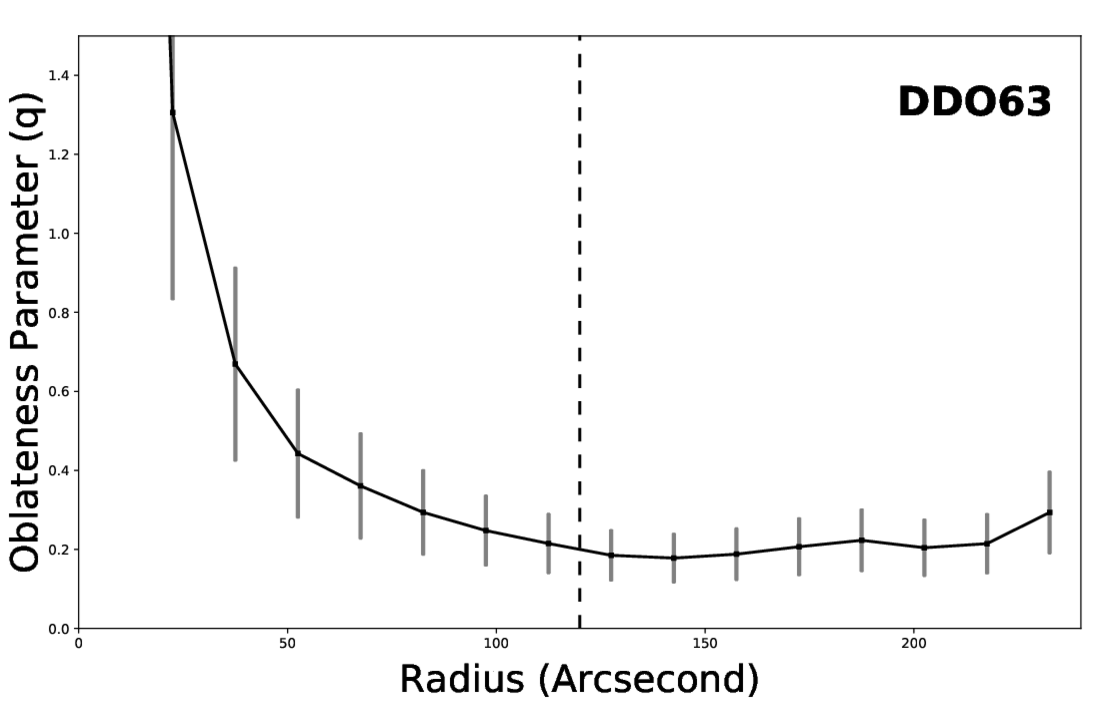}
   \includegraphics[scale=.1]{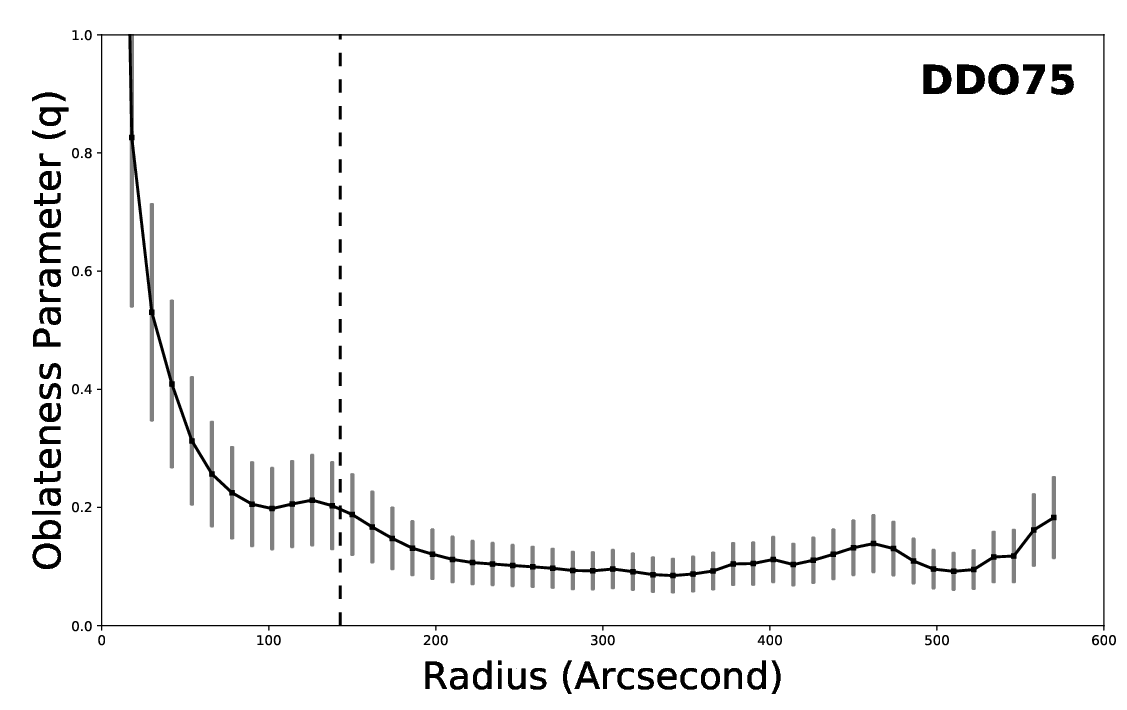}
   \includegraphics[scale=.1]{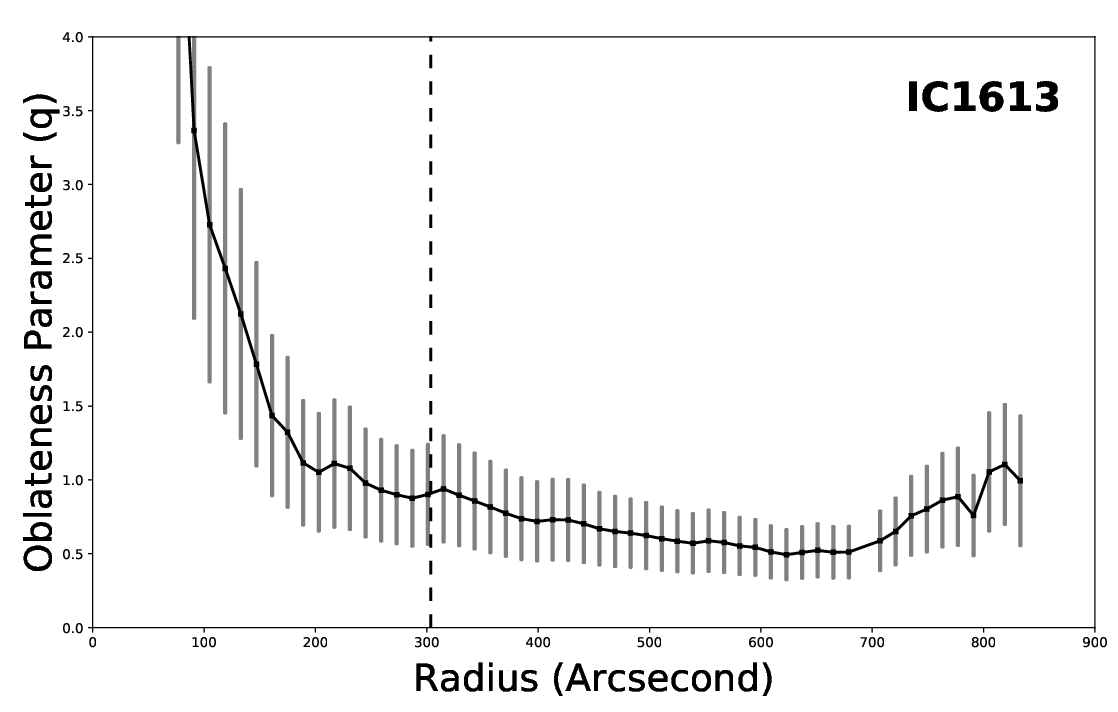}
  \caption{The top panel shows how $q$ varies with radius for three relatively massive galaxies whereas the bottom panel shows the same for less massive dwarf galaxies. The dashed vertical line marks the R$_{25}$ radius where the stellar disk ends.}
\end{figure}

\section{Our methods, galaxy sample and analysis}

We have assumed hydrostatic equilibrium for the gas disk in the vertical direction. The potential is composed of the stellar disk, molecular hydrogen gas ($H_2$), HI and dark matter. The dark matter halo has been modeled using the logarithmic potential; the stellar and gas disks have been assumed to have an exponential form. The detailed calculations are described in \citet{das.etal.2023}. The final expression used in the determination of the halo axis ratio $q=c/a$ is given by,

\begin{equation}
q^{2}~=~\frac{{v_{0}}^{2}{z_{g0}}^{2}}{R^2}\frac{1}{({\sigma_{\rm{zH\,{\textsc i}}}}^{2}~-~\pi Gz_{g0}\Sigma(HI)}    
\end{equation}

\noindent
where $v_0$ is the flat rotation velocity in units of km s$^{-1}$, the vertical disk scale height $z_{g0}$ and the radius are in kpc, the vertical HI velocity dispersion $\sigma_{\rm{zH\,{\textsc i}}}$ is in m s$^{-1}$, the gas surface density in $M_{\odot}pc^{-2}$ and the gravitational constant is $G=6.67\times10^{-8}$. The final expression after including all the constants becomes, 
\begin{equation}
q~=~\frac{{v_0}z_{g0}}{R}\times\frac{10^{3}}{[{{\sigma_{\rm{zH\,{\textsc i}}}}^{2}}~-~13.97\times10^{6}z_{g0}\Sigma(HI)]^{1/2}} 
\end{equation}
where $\Sigma(HI)$ includes the correction for helium in the HI gas surface density. 

We have applied this expression to a sample of 20 quiescent, nearly face-on galaxies, where the HI velocity dispersion over the cube represents $\sigma_{\rm{zH\,{\textsc i}}}$ and the HI surface density $\Sigma(HI)$ can be determined from the moment0 map.  The galaxy sample was derived from the The H I Nearby galaxy Survey (THINGS; Walter et al. 2008), and the Local Irregulars That Trace Luminosity Extremes; The H I Nearby galaxy Survey (LITTLE THINGS; Hunter et al. 2012) and two galaxies from the VLA ACS Nearby galaxy Survey Treasury (VLA ANGST) survey (Ott et al. 2012). Of the 20 galaxies, 8 had stellar masses $M(*)>10^{9}M_{\odot}$, and the remaining 12 galaxies are low mass dwarf galaxies. 

The disk flat rotation velocity was determined using the baryonic Tully-Fisher relation. The method for determining  H I velocity dispersion $\sigma_{\rm{zH\,{\textsc i}}}$ variation with radius is summarised in \citep{ianjamasimanana.etal.2017} and \citet{das.etal.2020}, and involves stacking the individual velocity profiles over radial bins in the data cube. This helps to have a good signal-to-noise ratio (S/N) for the $\sigma_{\rm{zH\,{\textsc i}}}$ values. The plots are shown in the Appendix of Das et al. 2023.

\begin{figure}[h]
\centering
  \includegraphics[scale=.20]{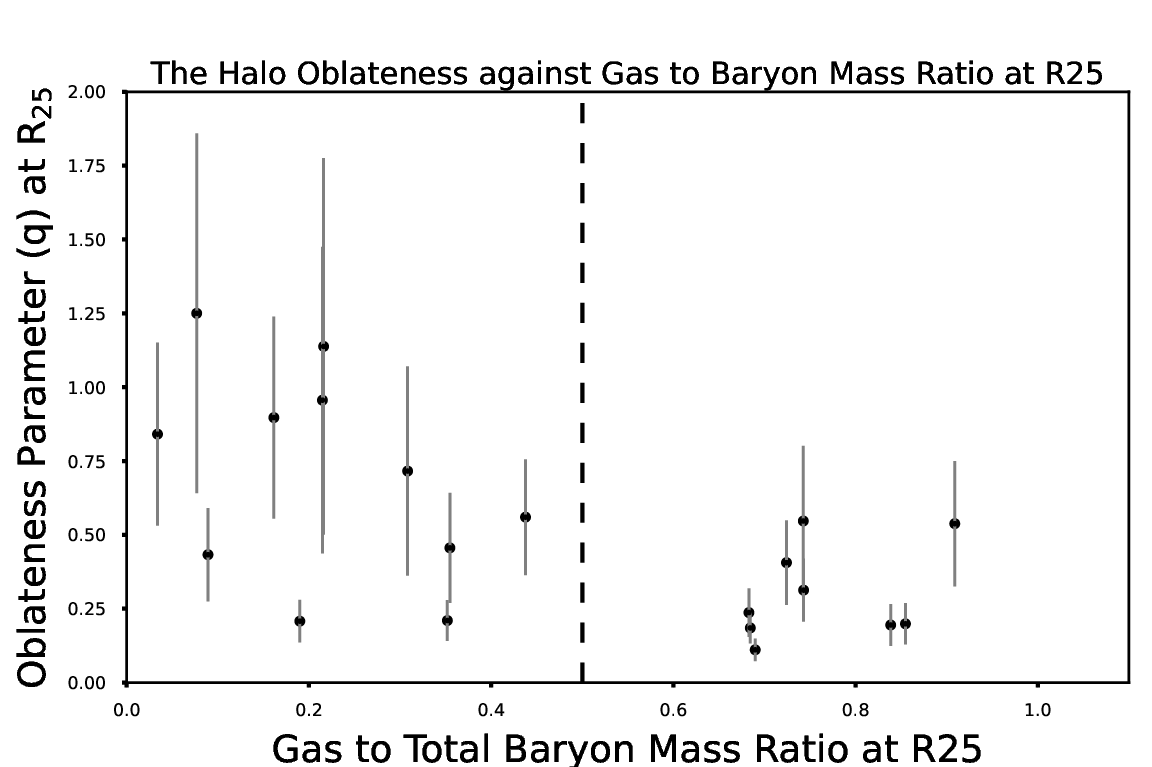}
  \caption{The above plot shows the oblateness parameter $q$ plotted against the gas to baryon mass fraction in galaxies. the dashed line is just the 0.5 mark, so that galaxies to the right of it are gas mass dominated and those to the left are stellar mass dominated. It is interesting to note that the gas dominated galaxies (which are all LSB dwarfs, have oblate halos.}
\end{figure}

\section{Main Results}
\noindent
{\bf 1.~}An important outcome of this study is that the $q$ does not vary much in the outer disks of galaxies at radii $R>R_{25}$ (see Figure 1). Of the sample galaxies, only the gas-rich, LSB dwarf galaxy DDO46 shows a significant decrease in q from q = 0.5 at R$_{25}$ to 0.3 at 2R$_{25}$. However, as shown in equation [1] and [2], $q$ depends on the thickness of the gas disk, $z_{g0}$, and this may vary with radius, especially in cases where the HI disk flares in the outer parts. We discuss this further in the discussion.\\
{\bf 2.~}The value of $q$ shows a significant dependence on galaxy mass, with $q$ increasing for increasing stellar mass. This suggests that the more massive galaxies with $M(*)>10^{9}M_{\odot}$ have rounder halos that are close to spherical and some even slightly prolate ($q\sim1.4$). However, there is some scatter and error; hence the Pearsons weighted correlation coefficient is 0.78. A larger sample will reveal whether this trend is real or not.\\
{\bf 3.~}Since a large fraction of our sample are gas dominated galaxies, we also plotted $q$ against the gas to baryon mass fraction (Figure 3). There is a clear trend for gas dominated galaxies to have lower $q$ values compared to the stellar dominated ones. All the gas dominated galaxies lying to the right of the dashed line in Figure 3 are LSB dwarf galaxies and have halo axis ratios  varying from $q=0.11$ to $0.55$. whereas the stellar dominated ones have a wide range of $q$ values, from $q=0.21$ to $1.25$.  

\section{Implications of our results}
\noindent
{\bf 1.~Halo Spin~:~}One of the most interesting results from this study is that gas dominated galaxies have more flattened or oblate dark matter halos compared to stellar dominated galaxies. Such galaxies are also dwarfs with low stellar masses. One reason could be that these galaxies have higher halo spin compared to massive galaxies. This has been suggested in previous studies \citep{kim.lee.2013} and is shown also shown by the modeling of the halo spin of the LSB dwarf galaxy UGC5288 by \citet{ansar.etal.2023}, and is discussed in a poster presentation by Sioree Ansar in this conference. The higher spin could be due to tidal torques experienced by the dwarfs at early epochs. Such galaxies are usually found in ioslated environments. On the other hand, the more massive stellar dominated galaxies may have experienced mergers and so they have a distribution of halo shape and spin. \\
{\bf 2.~Disk dynamical evolution~:~}One of the results of having oblate halos is that there is more halo dark matter associated with the disks. This can affect the disk dynamics. For example \citet{kumar.etal.2022} have shown that oblate halos delay bar formation in galaxy disks and consequently delay bar buckling. The formation and buckling of bars are important for triggering star formation and bulge growth, which are important parts of disk evolution. It is possible that spiral arm formation may also be curbed, as it is also a global disk instability like bars. The underlying reason for the slow growth of disk instabilities in oblate halos is that the disk becomes dynamically hotter. This is clearly noted in \citet{kumar.etal.2022}, where the velocity dispersion of the disk stars increases as the halo becomes more oblate.\\
{\bf 3.~The HI disk height~:~}In our study we have derived the HI gas disk height from the stellar disk radius using the ratio of $R_{d}/z_{g0}\sim0.85$ \citep{kregel.etal.2002}. This relation has been derived for stellar disks and may not be accurate for HI disks. This was adopted for simplicity, since there is no published study linking the disk scale length with HI thickness in galaxies. However, we do not think that including the flaring will change our basic result i.e. gas rich dwarfs have more oblate halos compared to massive, stellar dominated galaxies. To test this hypothesis we changed the disk thickness by 10\% for DDO187. For R=58.5$^{\prime\prime}$, the surface density of HI is 4.5 $M_{\odot} pc^{-2}$, and the HI dispersion is 12.5km/s (it has to be also deprojected with inc=39 degrees). Using the disk rotation in
equation [1] we find that $q$ changes from q=0.1109 to 0.1216. The change
in q is around 10\%. Hence, our overall result will still hold.\\
{\bf 4.~Flaring of HI disks and its effect on $q$~:~}In our study we have assumed that the disk thickness remains constant with radius. However, studies of edge on disk galaxies suggest that many HI disks start flaring i.e. the thickness increases beyond the $R_{25}$ radius \citep{obrien.etal.2010}. Looking at equation [1] and [2], this suggests that the $q$ increases with radius, or basically the halos become rounder with increasing radii in galaxies. 

\section{Conclusions}
\noindent
We present a method to determine the  oblateness $q$ of galaxy halos. Our method can be applied to gas-rich, face-on disk galaxies by applying the equation of hydrostatic equilibrium to the outer HI disks. We have used the HI velocity dispersion and HI surface densities of the disks to calculate $q$ at the $R_{25}$ and 1.5$R_{25}$ radii. Our study shows that q remains fairly constant in the outer disk regions, beyond the $R_{25}$ radius. Our results show that there is a significant correlation between the stellar mass and  $q$, which suggests that galaxies with massive stellar disks have a higher probability of having spherical or prolate halos, whereas the low-mass dwarfs have oblate halos. Our most significant finding is that gas-rich galaxies with M(gas)/M(baryons)$>$0.5 have oblate halos (q$<$0.55), but stellar mass dominated galaxies have a range of $q$ values. Also, the gas-dominated galaxies are dwarf galaxies. Hence, our results show that gas-rich dwarf galaxies have oblate halos whereas the larger galaxies have a range of halo shapes.  

\acknowledgements
M.D. acknowledges the support of the Science and Engineering Research Board (SERB) MATRICS grant MTR/2020/000266 for this research. RI acknowledges financial support from the grant CEX2021-001131-S funded by MCIN/AEI/ 10.13039/501100011033, from the grant IAA4SKA (Ref. R18-RT-3082) from the Economic Transformation, Industry, Knowledge and Universities Council of the Regional Government of Andalusia and the European Regional Development Fund from the European Union and financial support from the grant PID2021-123930OB-C21 funded by MCIN/AEI/10.13039/501100011033, by "ERDF A way of making Europe" and by the "European Union" and the Spanish Prototype of an SRC (SPSRC) service and support funded by the Spanish Ministry of Science and Innovation (MCIN), by the Regional Government of Andalusia and by the European Regional Development Fund (ERDF).

\begin{discussion}

\discuss{F. Lelli}{How will your results change if the HI disk is flared?}
\discuss{Answer} {A flared HI disk will have a larger vertical scaleheight and so the parameter $z_d$ will increase. If one examines the relation for $q$, it is clear that the denominator will decrease when $z_d$ increases, thus making the $q$ increase. So the halo will become rounder at larger and larger radii when the HI disk flares. }
\discuss{Gerhard Hensler}{We know that galaxies are affected by their CGM, e.g. infall of HVCs. Could you include such perturbation into your modelling, i.e. relaxing the equlibrium assumption?}
\discuss{Answer} {That is an interesting idea. The model in its present form assumes that the disk is in vertical hydrostatic equilibrium. So to include gas infall, such as that due to HVCs, we will have consider pertubation of the equilibrium equation. So it can be accommodated in the model but only for small rates of gas infall. }
\end{discussion}

\end{document}